\documentclass[11pt]{article}
\pdfoutput=1
\usepackage[margin=2cm]{geometry}
\usepackage{authblk}
\usepackage{mathrsfs}
\usepackage[colorlinks=true,citecolor=blue,linkcolor=blue]{hyperref}
\usepackage{amsmath,amssymb}
\usepackage{ifthen}
\usepackage{subcaption}
\usepackage{dsfont}
\usepackage{relsize}
\usepackage{slashed}
\usepackage{mathtools}
\usepackage{graphicx}             
\usepackage{url}
\usepackage{multirow}
\usepackage[dvipsnames]{xcolor}
\usepackage{csquotes}
\usepackage{nicefrac}
\usepackage{titletoc}
\usepackage{siunitx}
\usepackage[capitalize]{cleveref}
\usepackage{verbatim}
\usepackage[font={small}]{caption}

\newsavebox{\imagebox}

\newcommand{\E}{\mathrm{e}}

\newcommand{\GeV}{\text{GeV}}
\newcommand{\eV}{\text{eV}}
\newcommand{\dd}{\mathrm{d}}

\date{}
\title{Robustness of ARS Leptogenesis in Scalar Extensions}

\author[1]{Oliver Fischer\footnote{\href{mailto:oliver.fischer@liverpool.ac.uk}{oliver.fischer@liverpool.ac.uk}}}
\author[2]{Manfred Lindner\footnote{\href{mailto:lindner@mpi-hd.mpg.de}{lindner@mpi-hd.mpg.de}}}
\author[2]{Susan van der Woude\footnote{\href{mailto:susan@mpi-hd.mpg.de}{susan@mpi-hd.mpg.de}}}

\affil[1]{Department of Mathematical Sciences, University of Liverpool, Liverpool, L69 7ZL, UK}
\affil[2]{Max-Planck-Institut f\"ur Kernphysik, 69117~Heidelberg, Germany}

\begin{document}
 
\maketitle
 
\begin{abstract}
\noindent
Extensions of the Standard Model (SM) with sterile neutrinos are well motivated from the observed oscillations of the light neutrinos and they have shown to successfully explain the Baryon Asymmetry of the Universe (BAU) through, for instance, the so-called ARS leptogenesis. 
Sterile neutrinos can be added in minimal ways to the SM, but many theories exist where sterile neutrinos are not the only new fields.
Such theories often include scalar bosons, which brings about the possibility of further interactions between the sterile neutrinos and the SM.
In this paper we consider an extension of the SM with two sterile neutrinos and one scalar singlet particle and investigate the effect that an additional, thermalised, scalar has on the ARS leptogenesis mechanism. 
We show that in general the created asymmetry is reduced due to additional sterile neutrino production from scalar decays. 
When sterile neutrinos and scalars are discovered in the laboratory, our results will provide information on the applicability of the ARS leptogenesis mechanism.
\end{abstract}

\section{Introduction}
The discovery of the Higgs boson completed the Standard Model (SM) of particle physics \cite{ATLAS:2012yve,CMS:2012qbp}. However, there are still some loose ends which the SM is unable to connect. For instance the nature of Dark Matter (DM), the generation of neutrino masses or the observed Baryon Asymmetry of the Universe (BAU) are open questions which call for an extension of the SM. 
A very efficient theory framework that addresses these three questions simultaneously is the extension of the SM with three right-handed (or sterile) neutrinos.
Sterile neutrinos can be introduced explicitly in order to address the observations of neutrino oscillations, BAU, and DM in a minimal way. An excellent example is given by the so-called neutrino minimal Standard Model ($\nu$MSM) \cite{Asaka:2005pn, Asaka:2005an,Shaposhnikov:2008pf}, cf.\ also Ref.~\cite{Abazajian:2012ys} for a review.

It was shown by Akhmedov, Rubakov, and Smirnov that the existence of sterile neutrinos can explain the observed BAU via oscillations between the active and the sterile neutrino sectors \cite{Akhmedov:1998qx}, which is referred to as the ARS leptogenesis mechanism.
Within this mechanism CP violating interactions between sterile neutrinos and the active lepton sector result in a lepton number asymmetry in both the sterile and the active neutrino sector. 
Sphaleron processes will subsequently convert the lepton asymmetry stored in the (left-handed) active sector into a baryon asymmetry \cite{Klinkhamer:1984di} thus explaining the Baryon Asymmetry of the Universe. Note that this mechanism, unlike standard thermal leptogenesis \cite{Fukugita:1986hr} or resonant leptogenesis \cite{Pilaftsis:2003gt}, requires the sterile neutrinos to be non-thermal, i.e.~the interaction rate with the active leptons must be small. Consequently, ARS leptogenesis requires the sterile neutrinos to have GeV-scale masses. See for example Refs.~\cite{Canetti:2012kh,Drewes:2017zyw} for recent reviews and Refs.~\cite{Canetti:2010aw,Eijima:2018qke} for investigations into the available parameter space of such models.

In contrast to adding sterile neutrinos explicitly to the SM, their existence is motivated naturally in non-minimal extensions of the SM, for instance in models where an additional gauge symmetry is introduced. For example, when the $B-L$ numbers of the SM fermions are gauged, a $U(1)_{B-L}$ gauge factor is introduced, together with an additional gauge field and a scalar to make the gauge field massive via spontaneous symmetry breaking.
The existence of sterile neutrinos is required in this model to keep the theory anomaly-free \cite{Montero:2007cd}. In this case an explicit Majorana mass for the sterile neutrinos would be forbidden, however, it can be introduced dynamically through the same scalar that makes the gauge boson massive. 
Extensions of the minimal sterile-neutrino framework, and their effect on (ARS) leptogenesis, have been discussed in the context of many different theory frameworks:
in conformal models \cite{Khoze:2013oga,Khoze:2016zfi}; in $B-L$ extensions \cite{Caputo:2018zky}; in a Majoron model and axions \cite{Escudero:2021rfi}; in the context of inflatons \cite{Shaposhnikov:2006xi}.
Particularly interesting is the observation that resonant leptogenesis can be possible for heavy neutrinos as light as 500 GeV if their decays are assisted by additional scalars \cite{Alanne:2018brf}.

A smoking gun for a non-minimal neutrino sector would be the discovery of additional scalar resonances in the laboratory.
Scalar particles are searched for extensively at the LHC and excesses in recent data seem to point toward additional scalar degrees of freedom.
The CMS collaboration measured an excess in diphoton events \cite{CMS:2018cyk} that could be a scalar resonance at about 96 GeV, compatible with an excess in $b\bar b$ from LEP, cf.\ Refs.~\cite{Cao:2016uwt,Biekotter:2019kde}.
Moreover, there are excesses in multi-lepton final states pointing towards a heavy scalar with a mass of about 270 GeV \cite{vonBuddenbrock:2016rmr,vonBuddenbrock:2017gvy} that is connected to diphoton excesses in many signal channels pointing toward a resonance at 151 GeV \cite{Crivellin:2021ubm}. 
Last but not least, there are also some less significant excesses in four-lepton final states with invariant masses around 400 GeV and above \cite{ATLAS:2020zms}.
Clearly more data is needed to determine if any of these excesses will turn into a discovery.

In this paper we consider how an additional thermalised scalar would affect the efficiency of the ARS mechanism to address the BAU. Therefore we extend the SM with two right-handed neutrinos and a scalar boson, corresponding to an effective model that can in principle explain both neutrino oscillations and the observed BAU. This model can be interpreted as an extension of the so-called $\nu$MSM \cite{Asaka:2005an} or as a $B-L$ symmetric model where the gauge boson is many orders of magnitude heavier than the other SM extending fields.

This article is structured as follows. In \cref{chapter2} we will start with a discussion of the model including the sterile neutrinos and the additional scalar. Following this, thermal effects of the scalar on the dynamics of the sterile neutrino will be summarized. Afterwards, the kinetic equations, needed to calculate the lepton asymmetry, will be reviewed and the effect of an additional scalar will be discussed. \Cref{chapter2} will end with a discussion on the timescales which are relevant for successful leptogenesis via oscillations. \Cref{chapter3} will start with a general discussion on the available parameter space, where general arguments are used to determine which parameter ranges could lead to non-trivial dynamics. In the remainder of this section results from explicit calculations will be given. The results can be used to determine how the additional scalar affects ARS leptogenesis. The article will be wrapped up with our conclusions.

\section{Theory framework}
\label{chapter2}
We consider a minimal extension of the scalar sector with a real scalar singlet and a minimal extension of the fermion sector with two right-handed neutrinos (or, analogously, sterile neutrinos). 
The latter are motivated and constrained by the observed neutrino oscillations and we also consider LHC constraints on scalar resonances for the former, both of these constraints limit the model parameters at zero temperature. As we want to study early Universe cosmological implications of this framework, we also include finite temperature effects.

\subsection{The Model}
For concreteness we introduce $B-L$ symmetry with a corresponding $U(1)_{B-L}$ gauge factor that is spontaneously broken at an energy scale far above the electroweak scale.
In this model the field content beyond the SM is given by three additional sterile neutrinos $N_i,\,i=1,2,3$ and a complex scalar singlet $S$ that carries twice the charge of $N_i$ under the $B-L$ symmetry (typically lepton number $2$).
For the sake of minimality we make two assumptions:
the third sterile neutrino $N_3$ is decoupled and does not contribute to our discussion;
the gauge boson corresponding to the $B-L$ symmetry can be neglected, e.g.\ because its gauge couplings are sufficiently small or its mass is much larger than the other particle's masses. 
This leaves us with a real scalar boson $S$ and two sterile neutrinos.

In this scenario the following Yukawa terms can be added to the Lagrangian of the SM:
\begin{equation}
{\cal L}_Y = - F_{\alpha i} \bar{L}_\alpha H N_i - \frac{1}{2}Y_{ij} S \bar N^c_i N_j + (h.c.) \,.
\label{eq:lagrangianSNN}
\end{equation}
Above, $H$ is the SM Higgs field, $L_\alpha$ are the left-handed lepton doublets with $\alpha=e,\ \mu,\ \tau$ and $N_i$ with $i = 1, \ 2$ are the right-handed neutrinos that couple to $S$ with Yukawa-like coupling matrix $Y$. $F$ is a Yukawa-like coupling matrix describing the interactions between the right-handed neutrinos, the lepton doublet, and the Higgs boson. We work in a basis where the mass matrix of the sterile neutrinos is diagonalized, i.e.~the Yukawa matrix $Y$ is a diagonal matrix.
We remark at this point that in this model the lightest active neutrino is exactly massless due to the decoupled third sterile neutrino $N_3$.

The scalar potential in our model can be expressed as
\begin{equation}
V(S,H) = -\frac{1}{2} \mu_S^2 S^2 - \mu_H^2 H^\dagger H + \frac{1}{4} \lambda_S S^4 + \lambda_H (H^\dagger H)^2 + \frac{1}{2}\lambda_{SH} H^\dagger H S^2 \,,
\label{eq:scalarpotential}
\end{equation}
where the $\mu_i$ are mass parameters and the $\lambda_i$ are coupling constants of the scalar fields $S$ and $H$, where the former is a real scalar field and the latter is the complex isospin doublet of the SM Higgs boson.
Notice that terms that are odd in $S$ can be neglected because of the $B-L$ symmetry.

The scalars $S$ and $H$ can develop non-zero vacuum expectation values (vevs) when $\mu_i^2 > 0$:
\begin{equation}
\langle S \rangle = v_S^0, \qquad \langle H \rangle = \begin{pmatrix} 0 \\ \frac{1}{\sqrt{2}} v_{EW} \end{pmatrix}\,,
\end{equation}
As long as the mixing between the new scalar and the Higgs is small, the physical Higgs is dominated by the neutral component of the doublet $H$, with $v_{EW} = 246$~GeV.
As $H$ has isospin and hypercharge, this leads to spontaneous breaking of the electroweak symmetry as in the SM.

\subsection{Parametrisation}
\label{subsec:casasibarra}
The scalar sector contains two independent parameters that will be relevant for our discussion below: the scalar-Higgs coupling $\lambda_{SH}$ and the scalar self-coupling $\lambda_S$.
When the scalars $S$ and $H$ develop non-zero vevs, Dirac ($M_D$) and Majorana ($M_M$) mass matrices emerge for the neutrinos:
\begin{equation}
    M_D = F \cdot v_H,\qquad M_M = Y \cdot v_S^0 \,.
\end{equation}
In the type I seesaw approximation the small neutrino mass is given by $ M_D^2 / M_M$. Notice, however, that lepton number is broken in this setup only when the trace of $M_M$ is non-zero. Diagonalisation of the mass matrix yields the physical eigenstates, which are linear combinations of the interaction fields. 
We anticipate the requirement from leptogenesis for the sterile neutrinos to be quasi degenerate in masses,\footnote{The leptogenesis mechanism requires the mass splitting between the sterile neutrinos to be small, i.e.~the sterile neutrino masses to be highly degenerate. This is true for the ARS mechanism and also for resonant leptogenesis, see refs.~\cite{Klaric:2020phc,Klaric:2021cpi} for an extended discussion.} and introduce the parametrisation:
\begin{equation}
    M_{\pm} = M_N^0 (1 \pm \alpha)\,.
\end{equation}
The parameter $\alpha$ parametrises the mass difference of the two heavy neutrinos at zero temperature, and for $\alpha \ll 1$ we have $M_- = |Y_{11}| v_S^0 \simeq |Y_{22}| v_S^0 = M_+$ (in the basis where $M_M$ is diagonal) such that we can approximate the masses for both sterile neutrinos $M_\pm$ via the zero-temperature mass
\begin{equation}
    M_N^0 = Y v_S^0\,,
    \label{eq:Nbaremass}
\end{equation}
where we introduced the new parameter $Y= (|Y_{11}|+|Y_{22}|)/2$ which can be used instead of $M_N^0$.
This yields the two parameters $M_N^0$ (or $Y$) and $\alpha$, which are independent for $\alpha \ll 1$. 

\begin{table}
    \begin{center}
        $\begin{array}{c|c|c|c|c|c}
       m_1 &m_2 & m_3 & \sin^2\theta_{12} & \sin^2\theta_{13} & \sin^2\theta_{23}  \\
        \hline
    0 \ \eV & 8.68 \times 10^{-3} \ \eV & 5.03 \times 10^{-2} \ \eV  & 0.307 &  0.0218 & 0.545
        \end{array}$
    \end{center}
    \caption{Variables of the $\nu$MSM model, light neutrino observables from the Particle Data Group 2020 \cite{ParticleDataGroup:2020ssz}}
    \label{tab: variables}
\end{table}
The Yukawa coupling $F$ can be parametrised in a bottom-up and completely general way, based on observable low-energy data, the so-called Casas-Ibarra parametrisation for a $3+2$ neutrino sector \cite{Casas:2001sr}. 
For this parametrisation we use the following input parameters: the three neutrino mixing angles $\theta_{ij}$, the three active neutrino masses $m_i$, the two heavy neutrino mass eigenvalues (parametrised by $M_N^0$ and $\alpha$) and the four phases $\xi$, $\eta$, $\delta$ and $\omega$.
The mixing angles and active neutrino masses are known from neutrino experiments \cite{ParticleDataGroup:2020ssz}, see \cref{tab: variables}.

In order to compare our model to the ARS leptogenesis mechanism in the $\nu$MSM we fix the phases to the values used in Ref.~\cite{Asaka:2011wq}: $\xi = 1$,  $\omega = \pi/4 $, $\delta= 7 \pi /4$ and $\eta = \pi/3$. 
We remark that we checked that random variations of the internal parameters within the $1\sigma$ limits of the experimental measurements lead to ${\cal O}(1)$ modifications in entries of the Yukawa coupling matrix $F$.

In summary, in our model there are a total of five parameters that are free within certain limits, namely: $Y$, $\alpha$, $v_S^0$, $\lambda_S$ and $\lambda_{SH}$. We will discuss the constraints on these parameters below. 

\subsection{Constraints}
Here we list the considered zero-temperature constraints on the masses of the scalar from the LHC and on the mixing between active and sterile neutrinos. Limits from Early Universe cosmology on the sterile neutrino masses will also be discussed.

\paragraph*{Sterile neutrinos:}
Active-sterile neutrino mixing is constrained from precision measurements of the PMNS matrix, cf.\ Ref.~\cite{Antusch:2014woa}. Our use of the Casas-Ibarra parametrisation and the considered masses $M_N^0\leq 100$~GeV renders the resulting mixing parameters small compared to current limits.

For $M_N^0 < 1$~GeV, the decays of $N$ during the recombination period releases entropy into the thermal bath and impacts Big Bang Nucleosynthesis, which in general places strong limits on the mixing and mass parameters and requires in particular $M_N^0 \geq {\cal O}(0.1)$~GeV \cite{Canetti:2012kh}.

On the other hand, it has been shown that decaying heavy neutrinos with masses ${\cal O}(30)$~MeV \cite{Gelmini:2019deq} and sterile neutrinos that interact with additional scalars can alleviate the Hubble tension \cite{Fernandez-Martinez:2021ypo}. 
We will limit our discussion to $M_N^0 \geq 0.1$~GeV in the following.

\paragraph*{Scalar bosons:}
Additional scalar degrees of freedom that decay into pairs of gauge bosons have been searched for at the LHC, cf.\ e.g.\ the CMS report in Ref.~\cite{Roy:2021ooe}.
Non-observation restricts these particles to have masses above a few TeV with current data.
On the other hand, the recent LHC data includes excesses in the four-lepton invariant mass spectra that hint at additional resonances around 700 GeV \cite{Cea:2018tmm,Richard:2020jfd}. Even more convincing signals have been reported for some time now in non-resonant multi-lepton channels \cite{vonBuddenbrock:2019ajh} and recently in diphoton channels with associated production \cite{Crivellin:2021ubm}, which point at scalar bosons with masses around the electroweak scale. 
We therefore conclude that scalars with masses around the TeV scale are well motivated.

The measurement of the SM Higgs boson at 125 GeV limits its possible mixing with other scalar degrees of freedom. For the example of a single additional scalar resonance, this mixing can be constrained via precision measurements of the Higgs boson, and also with direct searches. Current constraints limit the sine of the mixing angle to ${\cal O}(0.1)$ \cite{Robens:2021rkl}.
The mixing angle and $\lambda_{SH}$ are related via
\begin{equation}
    \sin \alpha = \lambda_{SH} \frac{v_{EW} v_S^0}{M_S^2 - M_h^2}
\end{equation}
where $M_h = 125$~GeV is the mass of the observed Higgs boson.
If we assume that $M_h \ll M_S \simeq \sqrt{2 \lambda_S} v_S^0$, the limit on scalar mixing thus constrains
\begin{equation}
    \lambda_{SH} \leq 2 \times 0.1 \lambda_S\frac{v_S^0}{v_{EW}} \,,
    \label{eq:mixing}
\end{equation}
which implies that for $v_S^0 > {\cal O}(10) \times v_{EW}$ the interaction between the Higgs fields and $S$ can be strong, without affecting experimental constraints on the scalar-Higgs mixing. 
As we shall see below we will consider $v_S^0 \geq 10^6$, such that this limit can be met even if $\lambda_S \ll 1$.

\subsection{Finite Temperature effects}
In the early Universe both the scalar $S$ and the Higgs are in the symmetric phase, therefore, if $\mu_i^2 >0$, none of the particles have explicit mass terms. As discussed above, the scalars $S$ and $H$ can develop non-zero vacuum expectation values, which happens at a specific time in the early Universe, i.e.\ at $T_{EW} \simeq 140$~GeV for the Higgs boson, corresponding to the electroweak phase transition and sphaleron freezeout.
The $S$ symmetry breaking occurs at the temperature $T_S$.
For concreteness, we assume that this temperature is identical to the vev of $S$, i.e.\ $T_S = v_S^0$.
We implemented the time-dependent vev for $S$ through a numerical approximation of the Heaviside-theta function: 
\begin{equation}
    v_S(z) = v_S^0 \cdot \frac{1}{\E^{- 2 k (z- T_{EW}/v_S^0)} +1} \,,
    \label{eq:vs}
\end{equation}
with $k =10^5$. Furthermore, $z = T_{EW}/T$ is used as ``time'' variable. 
We remark that we do not implement a similar time-dependence for $v_{EW}$ as we only consider temperatures above $T_{EW}$, where $v_{EW}=0$.

We notice that for $z<1$ or, equivalently, $T> T_{EW}$, the Higgs field remains in the unbroken phase, there is thus no mixing between the two scalars. Any mixing induced after electroweak symmetry breaking does not affect ARS leptogenesis.

In general particles receive a thermal mass from their interactions with the thermal bath.  
In particular, the scalar $S$ can be thermalised via its interactions with the Higgs field for $\lambda_{SH}$ being sufficiently large, its thermal mass at one loop is given by 
\begin{equation}
    M_S(T)^2 = 2 \lambda_S (v_S(T))^2 +\frac{1}{4}\lambda_S T^2 + \frac{1}{6} \lambda_{SH} T^2\,,
    \label{eq:scalarmass}
\end{equation}
with $v_S(T)$ defined through \cref{eq:vs}. The first term corresponds to the zero temperature mass $M_S(T=0) = M_S^0$ in the limit where $\lambda_{SH}$ is negligible. In the following we shall approximate it as follows:
\begin{equation}
    M_S^0 = \sqrt{2\lambda_S} v_S^0\,.
    \label{eq:Sbaremass}
\end{equation}
This is an excellent approximation for $\lambda_S \gg {v_{EW} \over v_S} \lambda_{SH}$, and we require $v_S \geq 10^6$~GeV and $\lambda_{SH} \geq 10^{-4}$ as discussed below.

When the scalar $S$ is in equilibrium with the thermal bath the sterile neutrinos can also obtain a thermal mass from their Yukawa interactions with $S$ \cite{Khoze:2013oga}:
\begin{equation}
    (M_N^2(T))_{ii} =  (Y \cdot Y)_{ii} v_S(T)^2 + \frac{2}{3}\frac{1}{8} (Y \cdot Y)_{ii} T^2
    \label{eq:Nmass}
\end{equation}
for $i = 2, 3$. With $v_s(T)$ defined through \cref{eq:vs}.  
We remark that these thermal masses do not affect the Casas-Ibarra parametrisation, which is defined at zero temperature.

\subsection{Leptogenesis}

\begin{figure}
    \centering
    \begin{subfigure}{0.3\linewidth}
        \includegraphics[width=.8\linewidth]{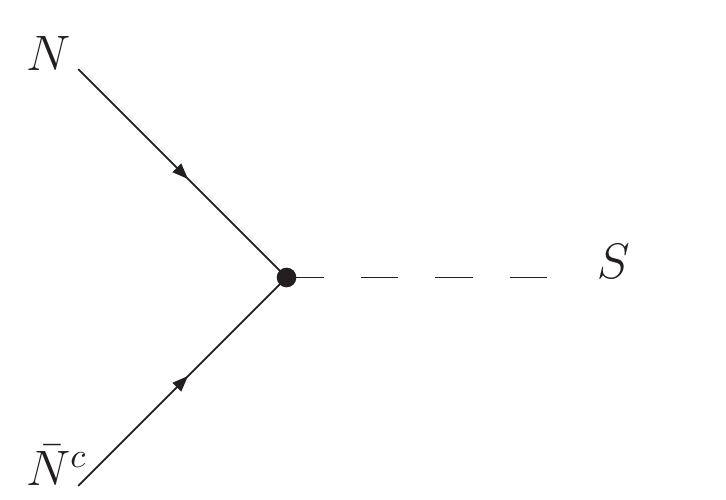}
        
        {\bf a}
    \end{subfigure}
    ~
    \begin{subfigure}{0.3\linewidth}
        \includegraphics[width=1\linewidth]{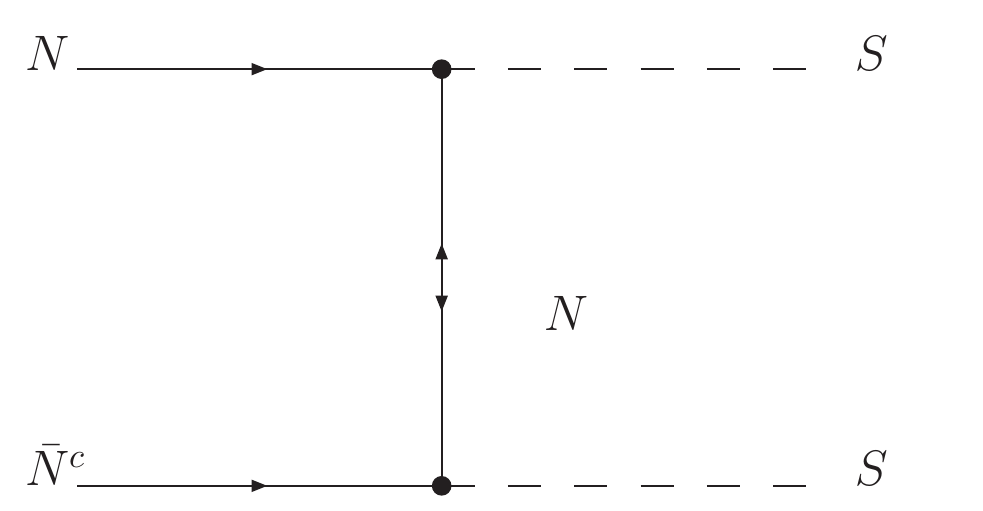}
        
        {\bf b}
    \end{subfigure}
    ~
    \begin{subfigure}{0.3\linewidth}
        \includegraphics[width=1\linewidth]{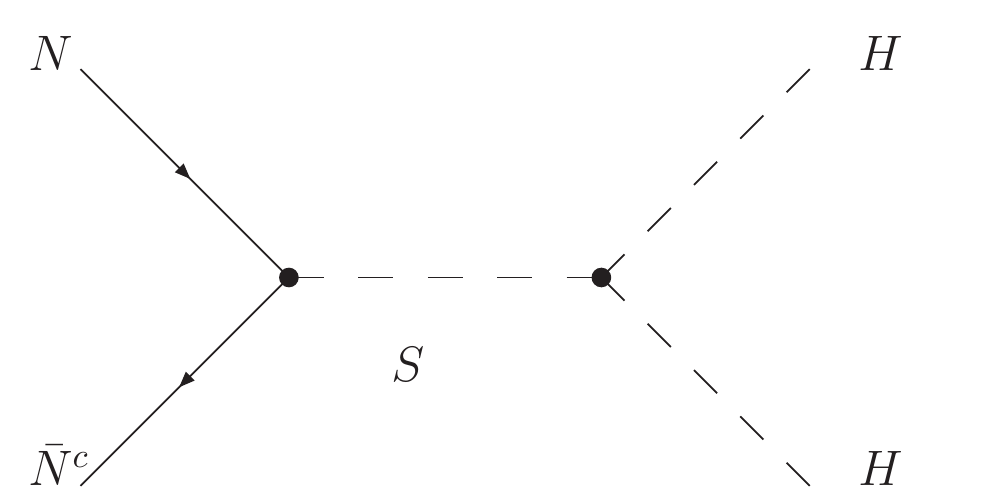}
        
        {\bf c}
    \end{subfigure}
    \caption{Processes equilibrating $N$ through $S$.}
    \label{fig:SNNprocesses}
\end{figure}

Leptogenesis relies on the existence of processes that fulfil the three Sakharov conditions: they must be out-of-equilibrium, they have to violate CP and they have to violate lepton number. The Sphaleron interactions then translate the lepton asymmetry in the active neutrino sector into a baryon asymmetry.
In models with 3 active neutrinos and 2 sterile neutrinos, a lepton asymmetry can be produced via oscillations between the active and the sterile neutrino sectors \cite{Asaka:2011wq}, where the dominant interaction is given by the Higgs boson-mediated process $N t \rightarrow L t$, with $L$ the lepton doublets, and $t$ the top quark.

The dynamics of the active leptons and right-handed neutrinos are determined via the kinetic equations, which describe the evolution of sterile neutrinos of each helicity $\rho_N$ and $\rho_{\bar{N}}$ as well as the evolution of the SM leptons. 
For convenience we consider the chemical potential $\mu_\alpha$ with lepton flavor $\alpha =e,\mu,\tau$, instead of the number densities for each particle and anti-particle species. 
The chemical potentials $\mu_\alpha$ depend on the specific momentum modes $x = k/T$, which makes solving them rigorously extremely difficult.

Fortunately, under the assumption that the sterile neutrino densities are proportional to the equilibrium density, i.e.~$\rho_N = R_N \cdot \rho_{eq}$ the kinetic equations 
can be simplified by taking the thermal average, corresponding to $k= 2 T$, without significantly affecting the numerical precision \cite{Asaka:2011wq}. Note that, like the full kinetic equations, this approximation also conserves the total lepton number, which we explicitly checked.

With these approximations the kinetic equations, in terms of $x = k/T$ and $z = T_{EW}/T$, can be written as \cite{Asaka:2011wq,Shuve:2014zua}:
\begin{align}
    \frac{d R_N}{d z} \frac{T_{EW}^2}{M_0 z}  = & - i [\langle H_N^0 \rangle + \langle V_N \rangle, R_N] -\frac{3}{2} \langle \gamma_N^d \rangle \{F^\dagger.F, R_N -1\} + 2 \langle \gamma_N^d \rangle F^\dagger.(A-1).F \nonumber  \\
   & -\frac{\langle \gamma_N^d \rangle}{2} \{F^\dagger.(A^{-1}-1).F, R_N\}  +\Gamma_S \\
    \frac{d\mu_{\alpha}}{d z} \frac{T_{EW}^2}{M_0 z}  = & -\gamma_\nu^d(T) [F.F^\dagger]_{\alpha \alpha} \tanh(\mu_\alpha) \nonumber \\ \nonumber
     & + \frac{\gamma_\nu^d(T)}{4} \left( \left( 1 + \frac{2}{\cosh(\mu_\alpha) }\right)  \left[ F.R_N. F^\dagger - F^*. R_{\bar{N}} .F^T \right]_{\alpha \alpha } \right.  \\ 
    & \left. - \tanh(\mu_\alpha)  \left[ F.R_N.F^\dagger - F^*. R_{\bar{N}} . F^T \right]_{\alpha \alpha } \right)
    \label{eq:kineticeq}
\end{align}
with $z=T_{EW}/T$, related to time and through the Hubble constant, $H = \frac{1}{2t} = \frac{T^2}{M_0}$, with $M_0 = 7.12 \times 10^{17} \GeV$. The time derivative is thus related to the $z$ derivative as; $\frac{\partial }{\partial t} = \frac{T_{EW}^2}{M_0 z} \frac{\partial}{\partial z}$. Definitions of $\gamma_N^d$ and $\gamma_\nu^d$, coming from the $Nt \rightarrow Lt$ interactions can be found in ref.~\cite{Asaka:2011wq}. Note that $\langle \ \rangle $ denotes the thermal averaging, for all terms $\sim 1/k$ this corresponds to $k \rightarrow 2T$ in the Maxwell-Boltzmann approximation. 
By taking the complex conjugate of the kinetic equation of $R_N$ the kinetic equation of $R_{\bar{N}}$ can be obtained straightforwardly.

The kinetic equations used in this paper, like the Boltzmann equations, are valid for relativistic systems close to equilibrium \cite{Lindner:2007am}. A full calculation requires the use of so-called Kadanoff-Baym equations, however, it was shown that in the context of thermal leptogenesis the Boltzmann equations are actually able to predict the lepton asymmetry relatively well, see e.g.~Ref.~\cite{Anisimov:2010dk}. Considering the uncertainty in the predicted BAU due to the other simplifications we have imposed on the kinetic equations, we deem it sufficient to use the kinetic equations as stated above to estimate the BAU. 

$H_N^0$ is the free Hamiltonian $\sqrt{k^2 +(M_N^0)_{ii}^2} \cdot \delta_{ij}$. $V_N$ is the effective potential and contains the medium effects. As mentioned before the tree level mass $M_N^0$ is defined as
\begin{equation}
    (M_N^0)_{ii} = Y_{ii} v_S(z)  \,,
\end{equation}
with $v_S(z)$ defined in \cref{eq:vs}. 
We remark that we are using Maxwell-Boltzmann statistics throughout, unless stated otherwise, thus $\rho_{eq} = \E^{-x}$.

$V_N$ is the effective potential of the sterile neutrinos, which describes the interaction with the plasma, is given by
\begin{equation}
    V_N = \frac{N_D T^2}{16 k} F^\dagger F + \frac{(M_N(T))^2}{2 k} =  \frac{N_D T^2}{16 k} F^\dagger F + \frac{2}{3}\frac{T^2}{16 k} Y\cdot Y \,,
\end{equation}
with $k=2T$ in the thermal averaged approximation. Whereas the first term comes from interactions of the sterile neutrino within the SM bath and is included with the $\nu$MSM formalism \cite{Asaka:2011wq}, the second term is due to interactions with the scalar.

The interaction terms in \cref{eq:lagrangianSNN} and \cref{eq:scalarpotential} introduce processes that connect the sterile neutrinos with the thermal bath, as shown in \cref{fig:SNNprocesses}. These are the scalar decay (and inverse decay) process {\bf (a)}, $t$-channel $N$-scalar boson scattering {\bf (b)}, and $s$-channel $N$-Higgs boson scattering {\bf (c)}.
We notice that the process {\bf (c)} occurs only after $S$ symmetry breaking and is proportional to the product $(Y \lambda_{SH})^2$ (and is further suppressed by a factor $(T/m_S)^4$ for $T< m_S$), while the process {\bf (b)} is proportional to $Y^4$. The decay process {\bf (a)} on the other hand is proportional to $Y^2$, which makes it the dominant process for $Y,\lambda_{SH} \ll 1$, such that we neglect the other terms in the following.

We remark that the $U(1)_{B-L}$ gauge boson brings about further interactions between the sterile neutrinos and the SM fermions. If the gauge boson is massless prior to $S$ symmetry breaking $N$ will be in thermal equilibrium at early times. In the following, we shall assume that the gauge boson has interaction rates that are sufficiently suppressed, for instance through a combination of tiny couplings or large gauge boson masses, such that $R_N=0$ for times that are early compared to the timescale where the ARS leptogenesis mechanism is efficient.

The process {\bf (a)} adds a new term to the kinetic equations, which corresponds to sterile neutrino production from the decays of the scalar $S$. While $S$ is thermalised with the SM particles, it can act as a source for $N$ and $\bar N$ production, via its decay \cite{Shaposhnikov:2006xi,Drewes:2015eoa}
\begin{equation}
   \Gamma_S = \frac{Y\cdot Y}{16 \pi} \frac{1}{\rho^{eq}(x)} \frac{M_S(z)^2}{T_{EW}}\frac{z}{x^2} \int_{y_0}^\infty n_s(y) \dd y\,,
    \label{eq:scalardecay}
\end{equation}
with $y_0 = x+\frac{z^2}{4x}\frac{M_s^2}{T_{EW}^2}$ and $n_s(y) = \E^{-y}$, i.e.~also here the equilibrium density is approximated by the Maxwell-Boltzmann distribution.

We remark here that we consider only production of sterile neutrinos of momentum $k=2T$, which is fixed through the parameter $x$ in \cref{eq:scalardecay}. 
Sterile neutrino distributions from scalar decay that are not Boltzmann-like should lead to very similar results, since this is the most relevant momentum mode for the ARS mechanism.

The source term in \cref{eq:scalardecay} depends on the coupling parameters, $Y,\lambda_S,\lambda_{SH}$ and $v_S^0$ through the thermal mass $M_S(T)$ (or $M_S(z)$).
Notice that the same process contributes to $\rho_{\bar{N}}$, which is accounted for with a factor $1/2$, compared to the decay rate stated in ref.~\cite{Khoze:2013oga}. We remark that this term also acts as a sink for the sterile-neutrino sector through inverse decays, $\bar{N}^c N \to S$. However, in the following we consider sterile neutrinos to be out of equilibrium, such that inverse decay can be neglected. If the scalar-sterile neutrino interaction would equilibrate long before Sphaleron freeze-out the sterile neutrino sector washout would remove any produced asymmetry.

\subsection{Time-scales}
It is useful to consider the time scales for understanding the dynamics of leptogenesis \cite{Shuve:2014zua}. Within ARS leptogenesis there are several important timescales, as we discuss below. 

\paragraph*{Sphaleron freeze-out:} The possibly most important time scale is set by the Sphaleron freeze-out temperature, which happens around $T \sim T_{EW}$. In terms of our time variable $z$ this temperature corresponds to $z=1$. In order to have efficient Baryon Asymmetry production from the lepton asymmetry in the active sector, lepton asymmetry must be produced \textit{before} Sphaleron freeze-out. Any lepton asymmetry produced after Sphaleron freeze-out is irrelevant for the BAU. The total baryon asymmetry is given by
\begin{equation}
    Y_{\Delta B} = -\frac{28}{79} (Y_{\Delta L_e} +Y_{\Delta L_\mu}+Y_{\Delta L_\tau})\,,
\end{equation}
where the $Y_{\Delta \alpha}$ correspond to the asymmetries for leptons $\ell_\alpha$.

\paragraph*{$S$ symmetry breaking:} At the temperature $T_S$ the scalar $S$ develops its vev $v_S^0$, and as discussed above we assume for simplicity that $T_S = v_S^0$. This implies that at the time $z_S = T_{EW}/T_S = T_{EW}/v_S^0$ the sterile neutrinos and the scalar receive bare masses as defined in \cref{eq:Nbaremass} and \cref{eq:Sbaremass}, respectively.
We remark that $S$ can remain thermalised until $z=1$ as is discussed below.

\paragraph*{Oscillations:} Another relevant timescale is related to the oscillations within the sterile neutrino sector; $t_{osc}$. Due to small mass splitting between the two heavy sterile neutrinos each sterile neutrino propagates at a slightly different speed through the plasma; this results in a phase shift between the two sterile neutrinos, which is crucial for developing the lepton asymmetry. This timescale, defined by the time it takes to build up an  $\mathcal{O}(1)$ phase difference, can be defined as a function of $z$ as
\begin{equation}
    1 = \int_0^{t_{osc}} \frac{\Delta M^2}{4T}  \dd t \,,
    \label{eq:zosc}
\end{equation}
where we introduced the sterile neutrino thermal mass splitting:
\begin{equation}
    \Delta M^2 = |(M_{N_1}(T))^2 - (M_{N_2}(T))^2|\,.
    \label{eq:oscillationtime}
\end{equation}

We notice that the absolute sterile neutrino mass splitting depends on the temperature, such that $z_{osc}$ has to be evaluated via \cref{eq:zosc} as a time-dependent quantity.

We remark that for $z \gg z_{osc}$ the oscillations become increasingly fast, and solving the full differential equations becomes computationally expensive. Following \cite{Shuve:2014zua} we solve the full calculations up to $z = N z_{osc}$, and for $z_{osc} < z \leq 1$ only the diagonal parts of the differential equation are solved. 
This is done by setting all off-diagonal components of the right hand side of eq.~\eqref{eq:kineticeq} to zero at $z=z_{osc}$. 
The factor $N=20$ is chosen such that Baryon Asymmetry agrees within 0.5\% to the full calculations, as was explicitly checked for benchmark point A in \cref{tab:points}.

\section{Analysis and results}
\label{chapter3}
In the following we consider sterile neutrino masses below the electroweak scale, i.e. $M_N^0 \leq 100$~GeV.
Such masses are too small to allow for the standard thermal leptogenesis or resonant leptogenesis to produce the observed Baryon Asymmetry.
Instead, we will focus on the so-called ARS leptogenesis mechanism, where the asymmetry is produced through oscillations between the active and sterile neutrino sectors.

\subsection{Discussion of the parameter space}
Among the five model parameters, the limits in \cref{eq:mixing} allow for reasonably large mixing, e.g.\ $\lambda_{SH} \sim \lambda_S$ is allowed for $M_S \geq {\cal O}(1)$~TeV and $v_S^0\geq{\cal O}(10)$~TeV. We therefore consider $\lambda_S$ and $\lambda_{SH}$ to be free parameters.
The other three parameters, $\alpha$, $Y$, $v_S^0$, are subject to a number of constraints, following the considerations below.

\begin{figure}[!]
\centering
    \begin{subfigure}{0.3\textwidth}
        \subcaption*{$\alpha = 0.1$}        
        \includegraphics[width = \textwidth]{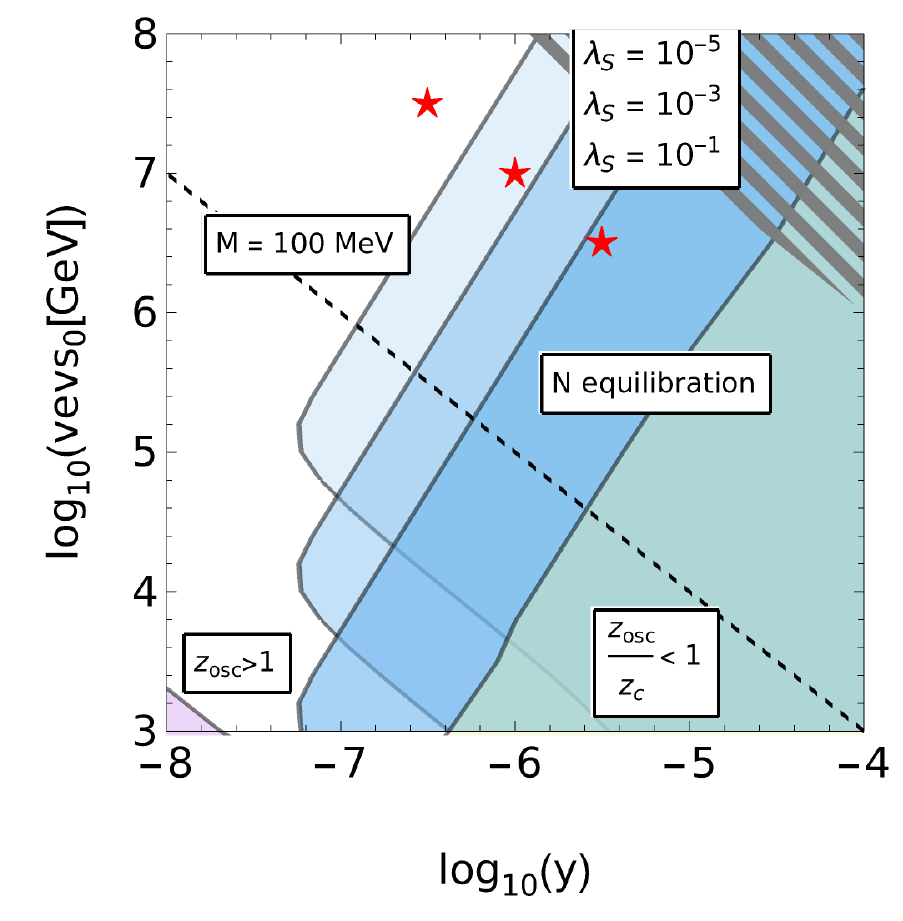}
    \end{subfigure}
    ~
    \begin{subfigure}{0.3\textwidth}
        \subcaption*{$\alpha = 10^{-3}$}
        \includegraphics[width=\textwidth]{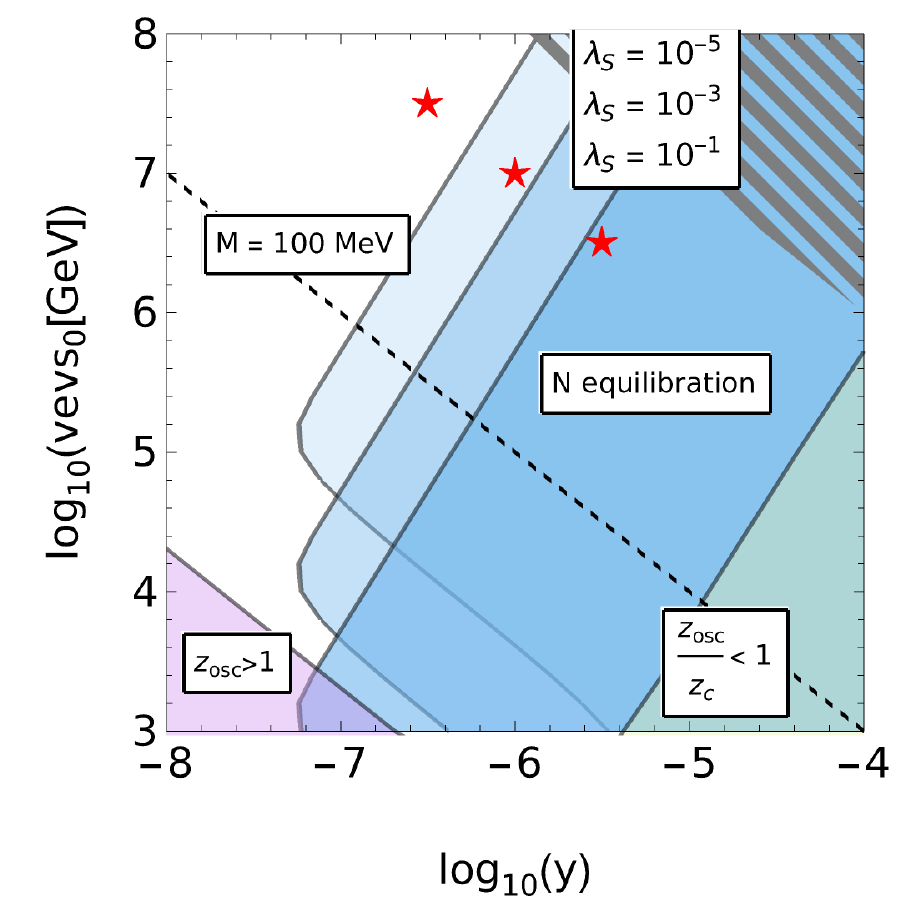}
    \end{subfigure}
    
     \begin{subfigure}{0.3\textwidth}
        \subcaption*{$\alpha = 10^{-8}$}
        \includegraphics[width=\textwidth]{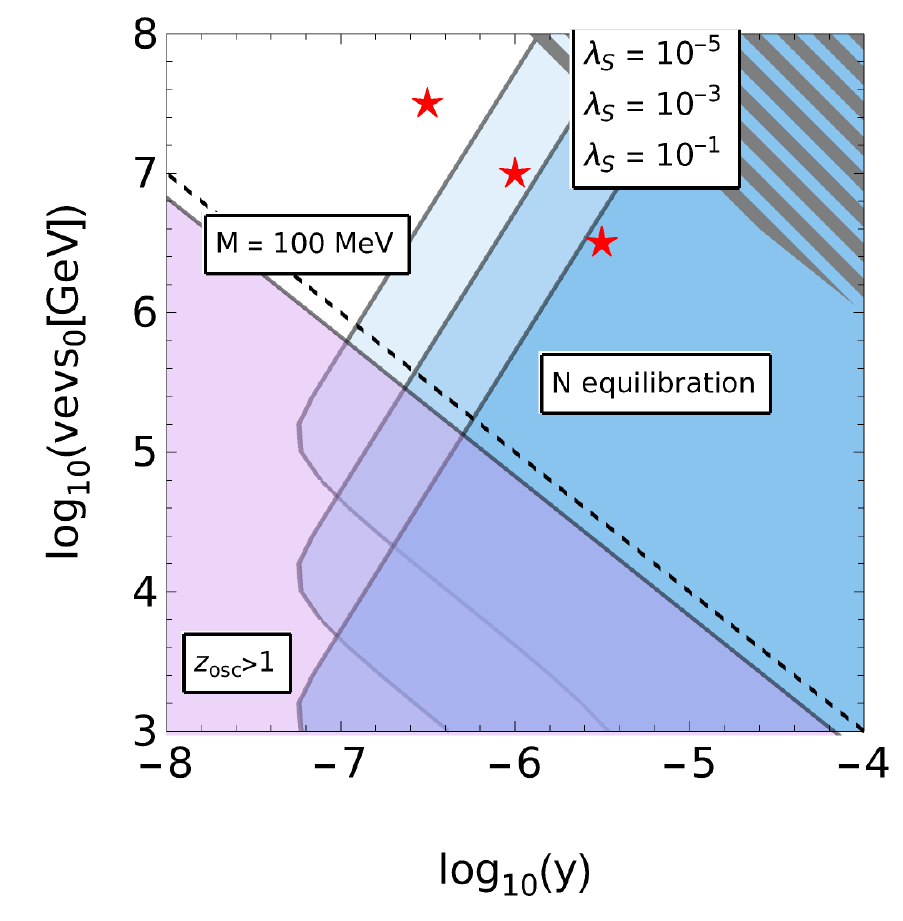}
    \end{subfigure}
    ~
    \begin{subfigure}{0.3\textwidth}
        \subcaption*{$\alpha = 10^{-8}$}
        \includegraphics[width=\textwidth]{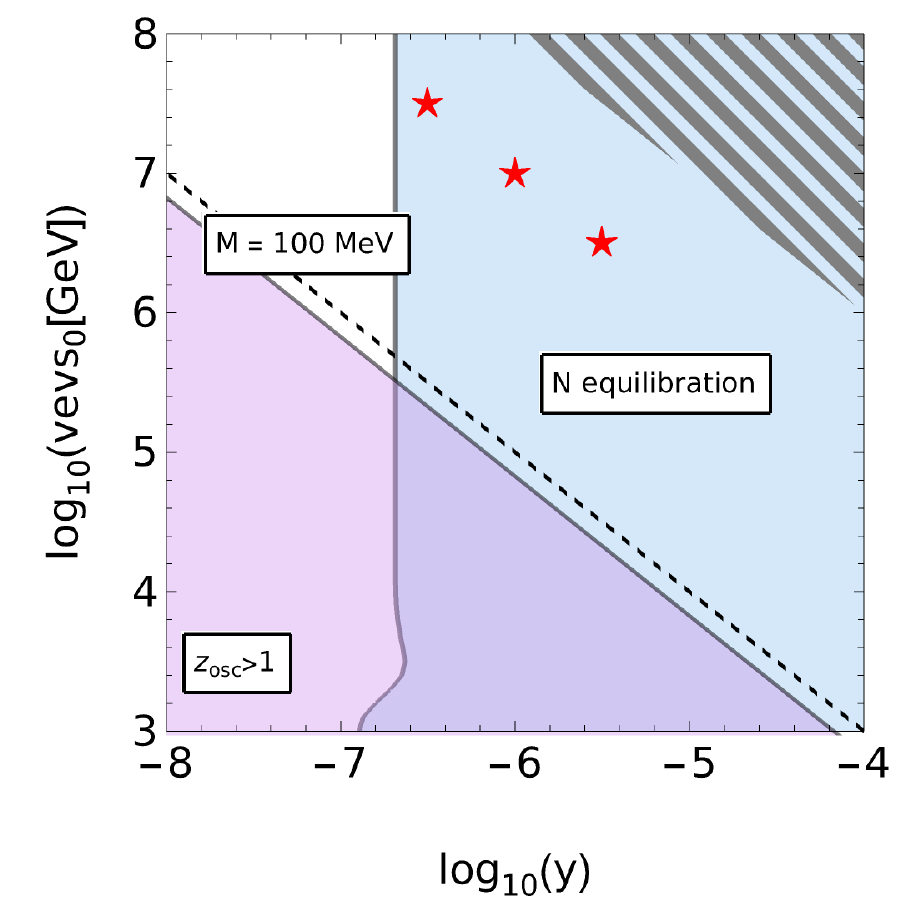}
    \end{subfigure}
    
    \caption{Parameter space with limits from successful leptogenesis as discussed in the text, in the projection $Y$ and $y_S$. Areas where the process {\bf (a)} ($S\rightarrow NN$) starts thermalising the sterile neutrinos are shown by the blue color, considering $\lambda_S = 10^{-1}, 10^{-3} \ 10^{-5}$. 
    Areas where the active-sterile oscillations occur after Sphaleron freeze-out are shown by the pink color. 
    The green area denotes where $z_S > z_{osc}$.
    The black hashed corner indicates sterile neutrino masses $M_N \geq 100$~GeV, and the dashed line corresponds to $M_N = 0.1$~GeV.
    Plotted are the benchmark points A, C and D, cf.\ \cref{tab:points}. For the lower right panel, $M_S^0=10$~TeV was fixed.
    Throughout, $\lambda_{SH} = 10^{-3}$ is fixed.}
    \label{fig:parameters}
\end{figure}

\paragraph*{Dominant scalar decay process:}
We remind ourselves that we assume the dominant $N$ interaction with the thermal bath to be given by process {\bf (a)}, $S\to \bar N N$. This process is only kinematically allowed if the induced thermal mass of the scalar is larger than the thermal mass of the sterile neutrinos, and in particular \cref{eq:scalardecay} is only correct, if $M_N\ll M_S$.
Therefore, our kinematic equations are valid if and only if $\sqrt{\lambda_s} \gg y$.
On the other hand, when the process $S \rightarrow N \bar{N}$ is kinematically forbidden one would have to consider the processes {\bf (b)} and {\bf (c)} of $N$-scalar scattering instead, cf.~\cref{fig:SNNprocesses}, which is beyond the scope of our discussion.

\paragraph*{Out-of-equilibrium $N$:}
In \cref{eq:scalardecay} we neglected the inverse decays, which implies that the sterile neutrinos have to have small number densities and thus be out of equilibrium. Explicitly, we chose the condition $\rho_N /\rho_{eq}<0.15$ at $z=1$. 
This choice is conservative: our numerical estimations, for some parameter choices, show that inverse decays are numerically negligible up to $R_N \sim 0.5$.
The two considerations above give conditions on both $v_S^0$ and $Y$, as functions of $\lambda_S$. These constraints are contained in the blue areas in \cref{fig:parameters}, for different values of $\lambda_S$ and fixed $\lambda_{SH}=10^{-3}$. 

\paragraph*{Early oscillations:}
As discussed above, active-sterile oscillations need to happen before Sphaleron freeze-out, such that the phase difference between the sterile neutrinos can create a Lepton asymmetry in the active sector that can be translated into a baryon asymmetry. Since Sphalerons freeze-out at $z=1$ the oscillations need to produce an order one phase shift before this time, which requires for the oscillation time: $z_{osc}<1$.
For different values of the mass splitting $\alpha$ this condition constrains on $Y$ and $v_S^0$ through the definition of the thermal mass in \cref{eq:Nmass}. The regions where oscillations are too slow are denoted by the pink areas in \cref{fig:parameters}.

\paragraph*{Relativistic $N$:}
Sterile neutrinos must remain relativistic up to $z=1$, such that the two helicity states of the sterile neutrino remain distinct.
Moreover, $N$ being relativistic also suppresses the amount of decays $N \rightarrow L H$, compared to the $2 \rightarrow 2$ interaction, thus validating neglecting this decay throughout. 
These considerations limit the sterile neutrino masses to $m_N \leq T_{EW}$. This implies that the black hashed area in \cref{fig:parameters} is nonphysical.

\paragraph*{Thermalised scalar:}
Our kinetic equations, as well as the decay rate into sterile neutrinos, make the implicit assumption that $S$ is in thermal equilibrium with the thermal bath for $v_S \geq T \geq T_{EW}$. For these temperatures the dominant interactions between $S$ and the SM is given by the $SSH^\dagger H$ term in \cref{eq:scalarpotential}.
We compute the interaction rate for the process $H^\dagger H \leftrightarrow SS$ as:
\begin{equation}
\Gamma = \sigma n(T)\,,
\end{equation}
where $n \propto T^3$ is the density of scalar bosons in the thermal plasma and $\sigma\propto \lambda_{SH}^2/T^2$ is the thermal cross section for this process. We evaluate the thermal cross section with the simplifying assumptions of massless Higgs bosons, and all external scalars having energies $E=2T$. With this, and neglecting the finite mass $M_S$,\footnote{The interaction rate drops quickly for $T\leq M_S$ due to phase space suppression. This is accounted for in the definition of the $S$ decay rate into sterile neutrinos \cref{eq:scalardecay}.} the reaction rate is identical to the Hubble rate under the condition
\begin{equation}
    \lambda_{SH} > 2.4\cdot 10^{-7} \sqrt{\frac{T}{\text{GeV}}}\,.
    \label{eq:lambdaSHcondition}
\end{equation}
Since we know that relevant dynamics require $T$ not to be too much larger than $T_{osc}$, let us consider $T\leq 10^{4} \cdot T_{EW}$.
The assumption that $S$ is thermalised for $z \geq 10^{-4}$ thus yields the condition: $\lambda_{SH} \geq 3 \cdot 10^{-4}$. 
In the following we shall always consider values for $\lambda_{SH}$, such that the condition in \cref{eq:lambdaSHcondition} is met.

\paragraph*{Time of scalar symmetry breaking:}
We have to consider the ordering of the two times $z_S$ and $z_{osc}$.
The parameter choice $z_S > z_{osc}$ indicates that $S$ symmetry breaking occurs relatively late, and that the sterile neutrino dynamics are dominated by their thermal mass rather than a fixed mass as is the case in the $\nu$MSM. This area is shown by the green areas in the upper panels of \cref{fig:parameters}.
The parameter choice $z_S < z_{osc}$ is expected to be dynamically closer to the $\nu$MSM.

\subsection{Successful leptogenesis}
\begin{table}[]
    \centering
    \begin{tabular}{c|c|c|c|c|c|c}
         Points & $\alpha$ & $\langle S \rangle [\GeV]$ & $y$ & $\lambda_S$ & $Y_{\Delta B}$  & remarks \\ \hline
         A & $10^{-8}$  & $10^{7.5}$ & $10^{-6.5}$ & $10^{-2}$ & $5.04\times 10^{-11}$& equivalent to $\nu$MSM\\ \hline
         B & $10^{-1}$ & $10^{3.5}$ & $10^{-6}$ & $10^{-5}$ &  $\sim 0 $& Within yellow area  \\ \hline
         C & $10^{-8}$  & $10^{7}$ & $10^{-6}$ & $10^{-2}$ & $4.96 \times 10^{-11}$ & relevant production of N\\ \hline
         D & $10^{-8}$  & $10^{6.5}$ & $10^{-5.5}$ & $10^{-2}$ & $ 1.46 \times 10^{-11}$ & large production of N \\ \hline
         E & $10^{-8}$ & $10^{8}$  & $10^{-6.5}$ & $10^{-9}$ & $ 2.5 \times 10^{-10}$ & ``enhancement''\\ \hline
    \end{tabular}
    \caption{
    Considered parameter space points $A,B,C \text{ and } D$. The parameters are the relative mass splitting $\alpha$, the sterile neutrino Yukawa coupling $Y$, the zero-temperature vev of the scalar singlet $v_S^0$, the scalar singlet self coupling $\lambda_S$. Given are the produced Baryon Asymmetry of the Universe $Y_{\Delta B}$ for each point, where the scalar-Higgs coupling $\lambda_{SH} = 10^{-3}$ has been fixed.
    }
    \label{tab:points}
\end{table}
It is important to realise that the process {\bf (a)}, cf.~\cref{eq:scalardecay}, creates sterile neutrinos and sterile anti-neutrinos in equal numbers and therefore by itself does not produce any asymmetry in the sterile sector. 
However, this process acts as a source for sterile neutrinos and thus increases $R_N$ and $R_{\bar{N}}$, which affects the lepton asymmetry production in the active sector via the kinetic equations.
For the discussion below we define a number of benchmark parameter points, listed in tab.~\ref{tab:points}, that correspond to different parameter space regions where successful leptogenesis is possible in principle.

\paragraph*{A: The limit of the $\nu$MSM:}
First, we consider the kinetic equations in \cref{eq:kineticeq} only, and use a fixed mass for the sterile neutrinos $M_N^0 = 10$~GeV, $\alpha = 10^{-8}$, which corresponds to the case considered in Ref.~\cite{Asaka:2011wq}. Solving the kinetic equations the total baryon asymmetry with the initial conditions $R_N=0$, $R_{\bar{N}}=0$, $\mu_\alpha=0$, $R_N(z)$, $R_{\bar{N}}(z)$ and $\mu_\alpha(z)$, we find the value $Y_{\Delta B} = -\frac{28}{79} Y_{\Delta L_{tot}} = 5.05 \times 10^{-11} $. 
This value differs by a factor of about four from the results in \cite{Asaka:2011wq}, namely $Y_B = 2.73 \times 10^{-10}$, which we checked is due to the different set of neutrino parameters.

Next we consider the benchmark point A, as defined in \cref{tab:points}, with the choice of $T_S \gg T_{EW}$. The small Yukawa coupling $Y$ makes the thermal contributions to the sterile neutrino mass negligible for $z\sim z_{osc}$, compared to its vev-induced mass of 10 GeV.
This benchmark point corresponds to the limiting case where the scalar interactions are negligible, and indeed the resulting asymmetry is identical to the one evaluated above.

\paragraph*{B: Late $S$ symmetry breaking:}
The case where the scalar $S$ develops its vev after the onset of neutrino oscillations defines $z_S > z_{osc}$. 
This, combined with the above discussed conditions show that only benchmark points with large relative mass splitting, small $v_S^0$, and relatively large Yukawa couplings can at least in principle generate an asymmetry, cf.\ the left panel of \cref{fig:parameters}. These parameters result in small sterile neutrino masses after the electroweak symmetry breaking, which in turn suppresses the magnitude of the Yukawa matrix $F$. 
As analytic estimates from Ref.~\cite{Akhmedov:1998qx} make us expect, the resulting baryon asymmetry from benchmark point B is consistent with zero, within computational uncertainties, cf.~\cref{tab:points}.

\paragraph*{C, D: Early $S$ symmetry breaking:}
Early breaking of the symmetry related to $S$ implies $z_S <z_{osc}$. 
In this regime the sterile neutrino mass is generally dominated by the zero temperature mass, i.e.\ it is temperature independent in very good approximation, and the oscillations are controlled by $M_N^0 = Y v_S^0$ for $T<v_S^0$. The dynamics are very similar to that of the $\nu$MSM, except for the additional $N$ production via the process {\bf (a)}.
The boundary of the blue area in the four panels of \cref{fig:parameters} indicates where $N$ production from $S$ decays increases the abundance of sterile neutrinos in the thermal bath to the point, where inverse decays become relevant. 
Parameter space points that are in the white area and close to the boundary with the blue area are expected to have enhanced production of sterile neutrinos.

The benchmark point C with $M_N^0 = 10$~GeV is close to this boundary for $\lambda_S = 10^{-3}$ and we notice that the resulting asymmetry is slightly reduced, compared to the result from benchmark point A. 
For comparison we also show the benchmark point D, which is inside the blue area and has a further reduced asymmetry compared to C.
Notice that, strictly speaking, the point D violates our assumption that the sterile neutrino densities are negligible. Estimates for the predicted BAU when inverse decay is included show that for moderate sterile neutrino production the suppression of lepton asymmetry production is actually reduced. This is a consequence of the reduced sterile neutrino abundance due to the inclusion of inverse decays. However, to fully understand the dynamics in the blue region more precise calculations are required, ideally including the momentum dependence or for example including more production channels, which is beyond the scope of this work.

\subsection{The effect of enhanced $N$ production}
We noticed above that leptogenesis can be successful only in the white regions of the parameter space, and that quantitative differences to the $\nu$MSM are to be expected only when $N$ production is not negligible, on the other hand, we expect that too large $N$ production will suppress the asymmetry production. Therefore we inspect the parameter space points that are at the boundary between the white and the blue area in \cref{fig:parameters} more closely.

\begin{figure}
    \centering
    \begin{subfigure}{0.3\textwidth}
         \includegraphics[width = \linewidth]{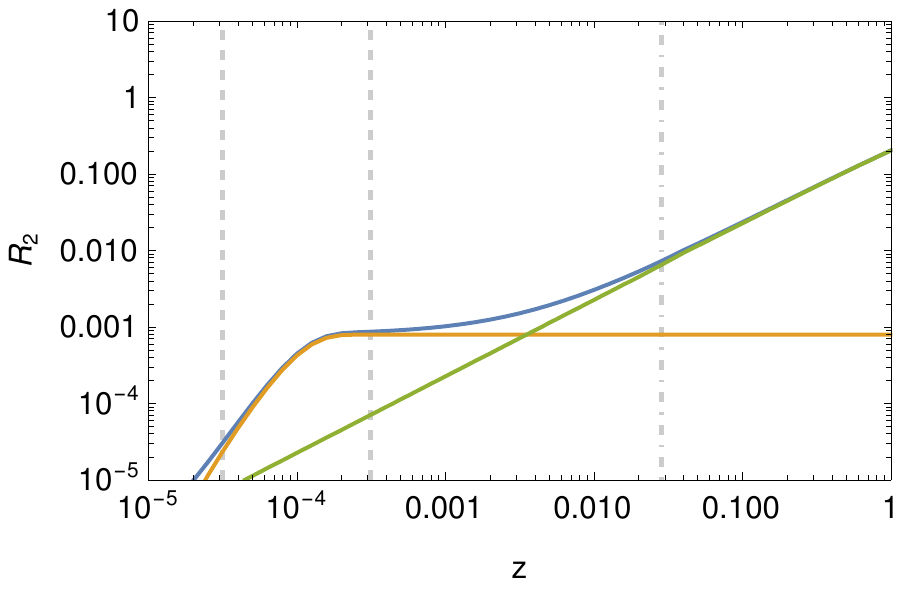}
         \subcaption*{A}
    \end{subfigure}
    ~
    \begin{subfigure}{0.3\textwidth}
         \includegraphics[width = \linewidth]{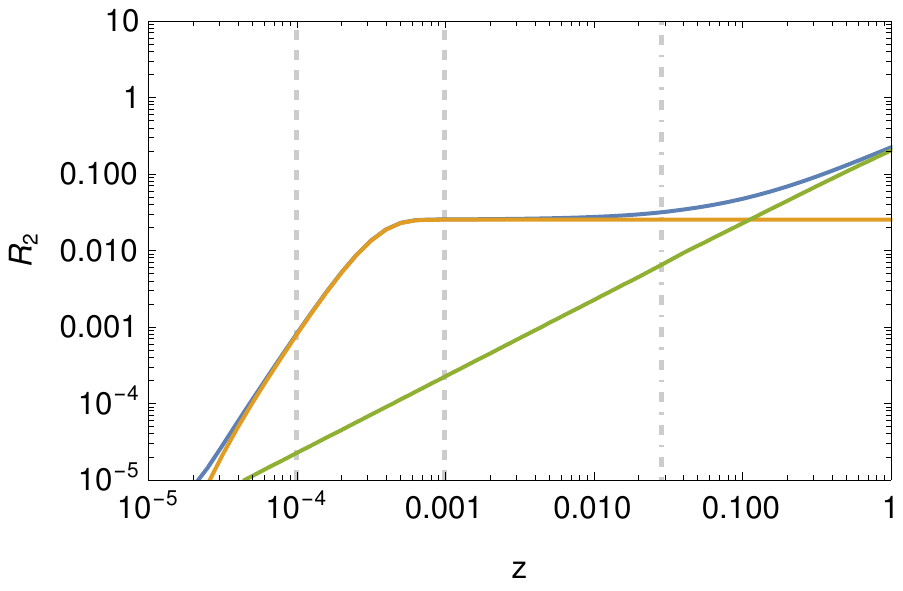}
         \subcaption*{C}
    \end{subfigure}
    ~    
    \begin{subfigure}{0.3\textwidth}
         \includegraphics[width = \linewidth]{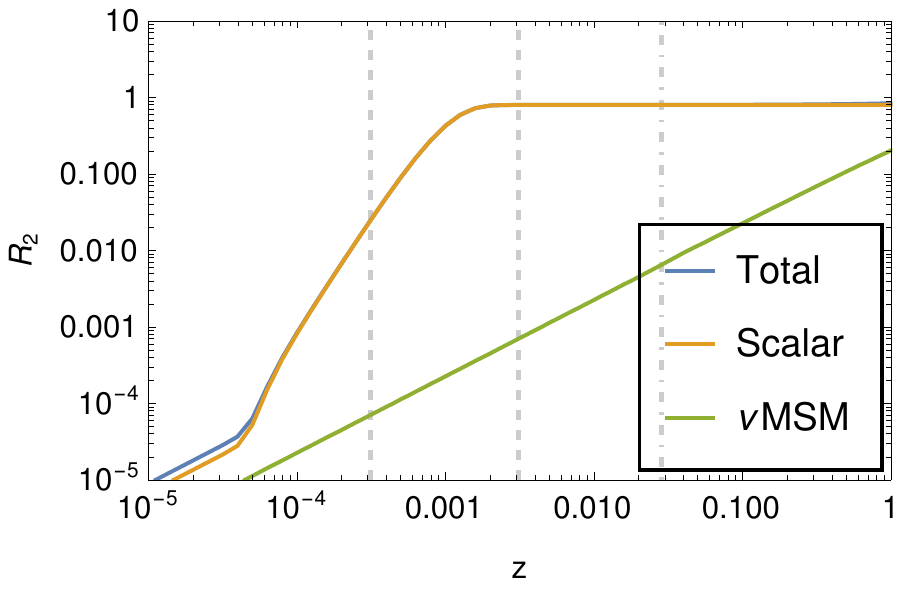}
         \subcaption*{D}
    \end{subfigure}
    \caption{Production of the sterile neutrino density $R_N$ as a function of the time parameter $z$ for the benchmark points A, C, D. The orange and blue line denotes total $N$ production and $N$ production via process {\bf a}, respectively. The green line denotes sterile neutrino production from $\nu$MSM dynamics. The vertical dashed lines indicate the time where $S$ decays are relevant, the dashed-dotted lines correspond to $z=z_{osc} = 0.028$. }
    \label{fig:RNprod}
\end{figure}

\paragraph*{Evolution of $R_N$:}
The evolution of the sterile neutrino density $R_N$ with $z$ for the benchmark points A, C and D, is shown in \cref{fig:RNprod}, wherein the blue line denotes production only via $S$ decays while the orange line includes the complete kinetic equations.
(Notice that the evolution of $R_{\bar N}$ is almost identical, apart from phase differences and from the relatively tiny difference that makes the asymmetry parameters.)
The figure shows that for the benchmark points A and C the sterile neutrino density $R_N$ remains below the equilibration limit of 0.15 at $z=1$. The point D, however, reaches equilibration for $z\simeq 10^{-3}$, which renders its result unphysical as the inverse decays have been neglected.
We observe that the main production of sterile neutrinos through scalar decay occurs for $T \sim \mathcal{O}(0.1) M_s(z)$, this region is denoted by the dashed grey lines in plots. For comparison the oscillation timescale is also shown in the plots as the dash-dotted lines.

\begin{figure}
    \centering
    \begin{subfigure}{0.45 \linewidth}
        \includegraphics[width= \linewidth]{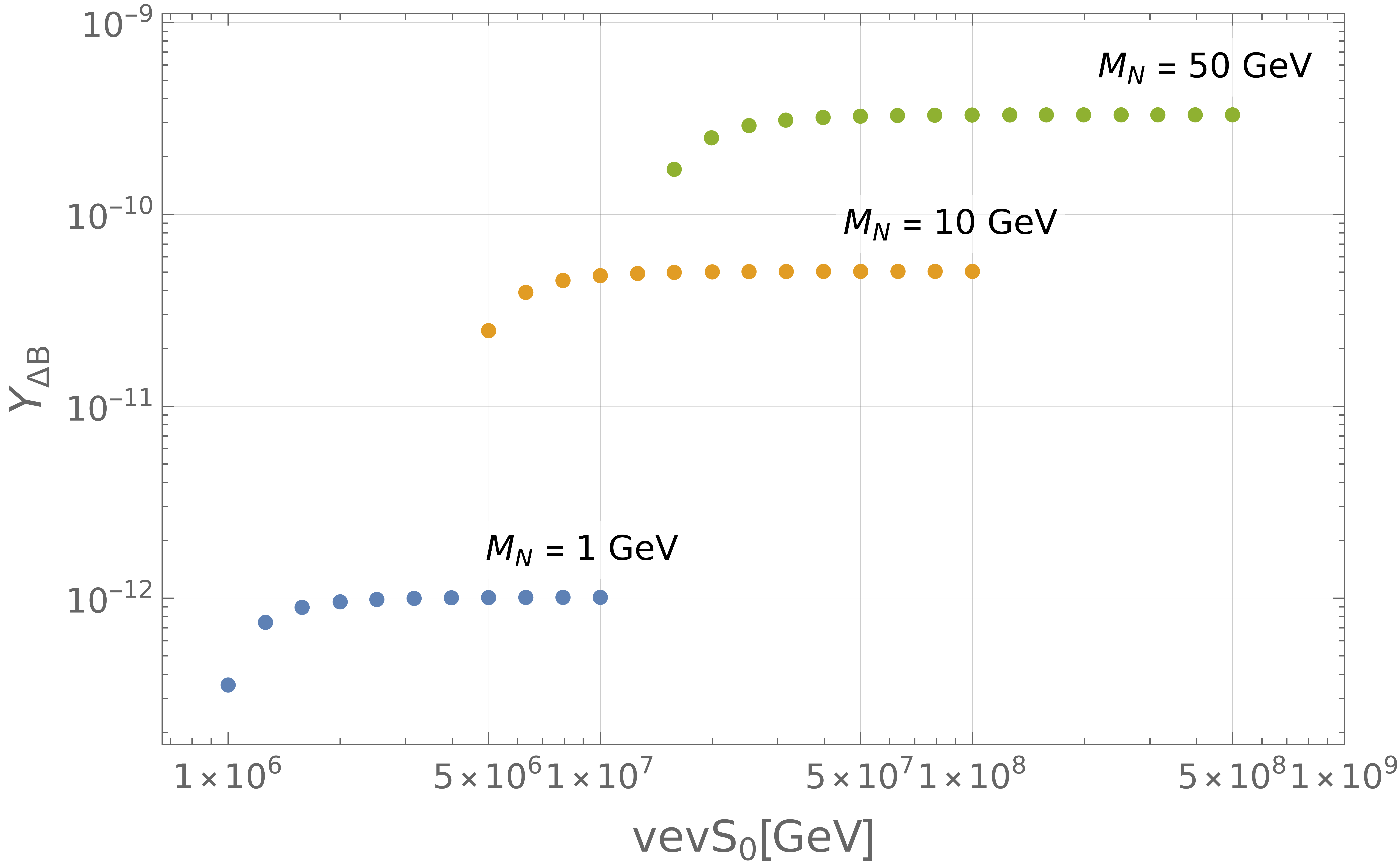}
    \end{subfigure}
    \begin{subfigure}{0.45 \linewidth}
        \includegraphics[width= \linewidth]{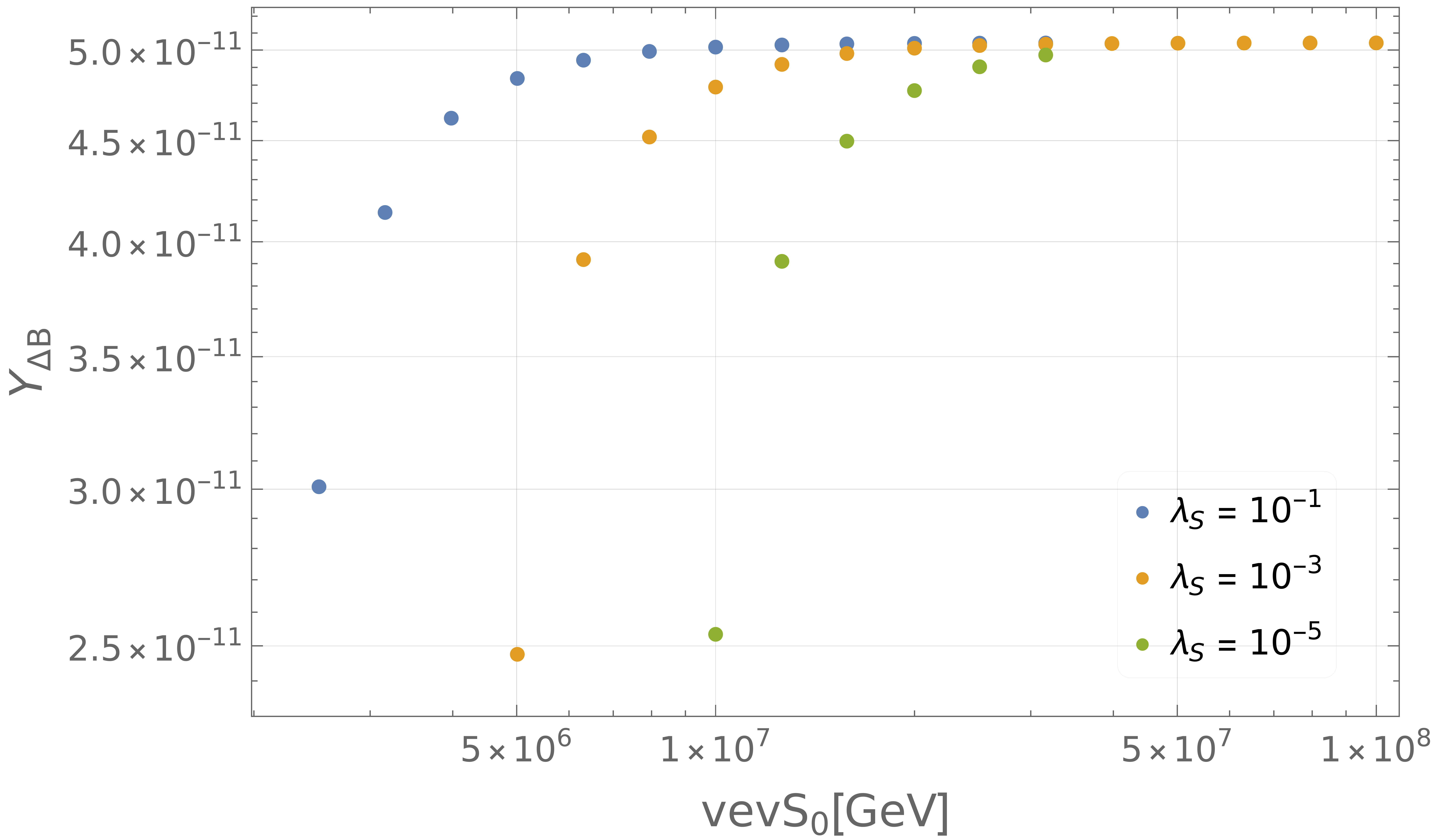}
    \end{subfigure}
    \caption{Baryon asymmetry production as a function of the $S$ vacuum expectation value $v_S^0$. {\it Left:} The three lines correspond to fixed sterile neutrino masses $M_N^0=1,10,50$~GeV. The $S$ self coupling has been fixed to $\lambda_S = 10^{-3}$.
    {\it Right:} The three lines correspond to $S$ self couplings $\lambda_S = 10^{-5}, 10^{-3}, 10^{-1} $. The sterile-neutrino mass has been fixed to $M_N^0=10$~GeV.
    For this figure $\lambda_{SH} = 10^{-3}$ has been fixed.}
    \label{fig:scans}
\end{figure}

\begin{figure}
    \centering
    \begin{subfigure}{0.7 \linewidth}
        \includegraphics[width = \linewidth]{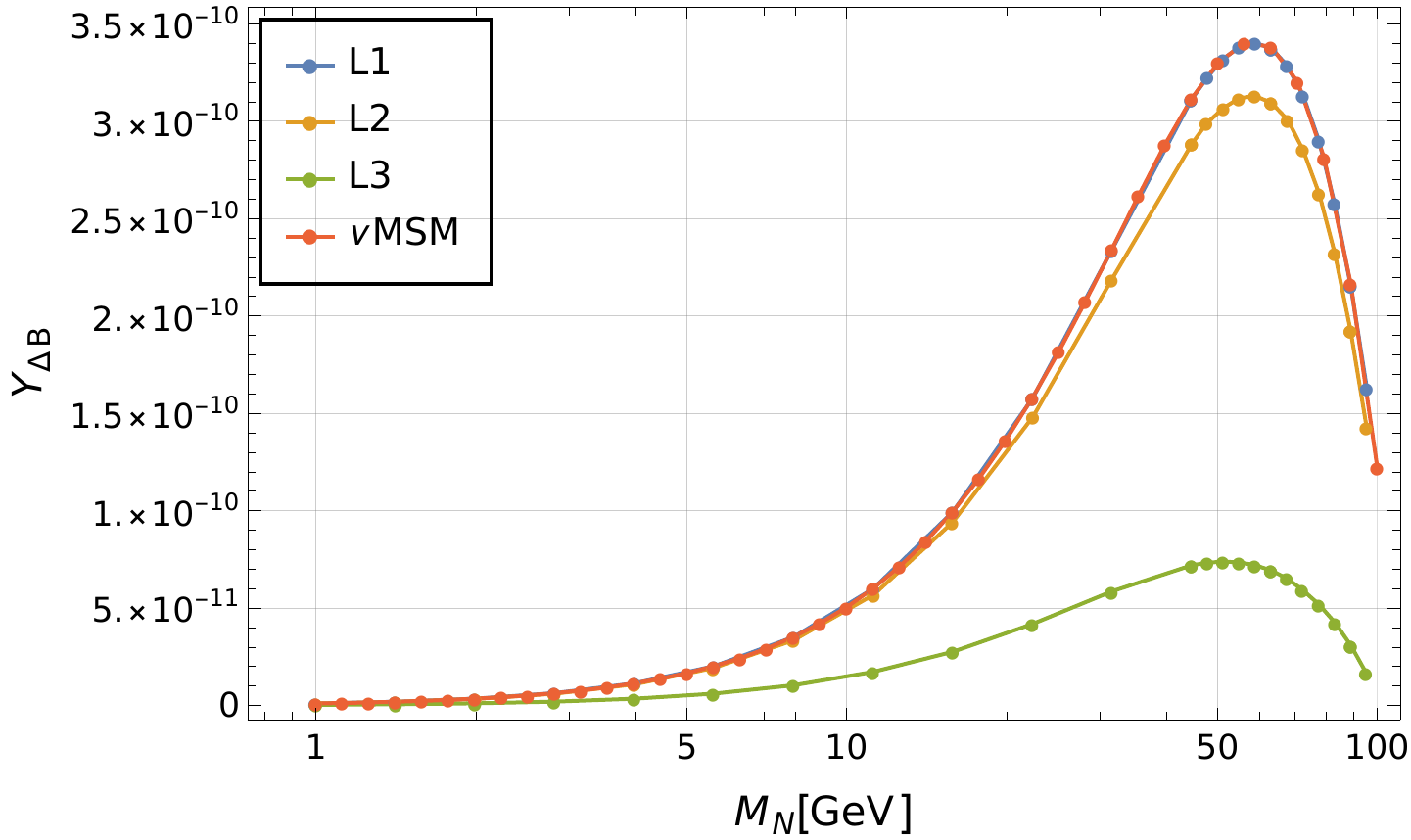}
    \end{subfigure}
    \caption{Total baryon asymmetry production as a function of the sterile neutrino mass $M_N^0$. The three lines $L_1,L_2,L_3$ are chosen parallel to the blue area. The remaining parameters are fixed to $\lambda_{SH}=10^{-3}$, $\lambda_S= 10^{-2}$ and $\alpha = 10^{-8}$. See text for more details.}
    \label{fig:Mndependence}
\end{figure}

\paragraph*{Varying the scalar vev:}
As discussed above, in the limit of large vev and early $S$ symmetry breaking, we reproduce the results of the $\nu$MSM.
Considering the effect of increased $N$ production, we keep the zero temperature neutrino mass $M_N^0$ fixed and vary $v_S^0$, which implies that $Y$ co-varies with the vev as $M_N^0/v_S^0$.

The left panel of \cref{fig:scans} shows three lines for fixed $M_N=1, 10, 50$~GeV, for each $M_N$ the self-coupling $\lambda_S=10^{-3}$ is fixed.
The right panel shows three lines for fixed $\lambda_{S}=10^{-5},10^{-3},10^{-1}$, and $M_N^0=10$~GeV is fixed.
In both panels $\lambda_{SH}=10^{-3}$ is used.
Both panels show that the asymmetry is reduced for smaller $v_S^0$.

The figure shows clearly how the asymmetry converges towards a fixed value when the interaction rate drops below some critical threshold.
Conversely the asymmetry drops with decreasing $v_S^0$ due to increased washout from additional sterile neutrino production through $S$ decays for increasing $Y$. 
The onset of the asymmetry reduction depends on the value of $Y$, as shown in the left panel, as well as the scalar mass $M_S$, as shown in the right panel.

\paragraph*{Varying the sterile neutrino mass:}
Here we consider the effect of varying sterile neutrino mass on the total baryon asymmetry production for different benchmark points.
Therefore we consider pairs of parameters $(v_S^0,Y)$ that are on a line parallel to the blue boundary in \cref{fig:parameters}.
We parametrise this line as
\begin{equation}
    \log(v_S^0) = 2 \log(Y) +L_i\,,
    \label{eq:parallel}
\end{equation}
where we fix $L_i=20, 18.5, 17.5$, which is, respectively, far away, close to, and inside the blue area for $\lambda_{S} = 10^{-2}$. 
We pick parameter points on these lines for 1~GeV~$\leq M_N^0\leq 100$~GeV, and we fix $\lambda_{SH} = 10^{-3}$ and $\alpha=10^{-8}$ for definiteness.

The resulting baryon asymmetry for each mass is shown in \cref{fig:Mndependence}, where the lines $L_1,L_2,L_3$ are denoted by the blue, orange, and green line, respectively, and where we show the result for the $\nu$MSM with the red line for comparison.
The figure shows clearly how the distance from the blue boundary determines the amount of washout and thus reduces the asymmetry production.

The enhancement for $M_N^0\sim 60$~GeV can be explained as follows: increasing $M_N^0$ also increases the Yukawa coupling between the sterile and active sector (through Casas-Ibarra parametrization), which increases the oscillation speed and therefore the asymmetry production. However, at some point the Yukawa coupling becomes so large that sterile neutrinos start to thermalise with the SM bath, and the resulting washout again reduces the produced asymmetry.

\subsection{Discussion}
\begin{figure}
    \centering
    \includegraphics{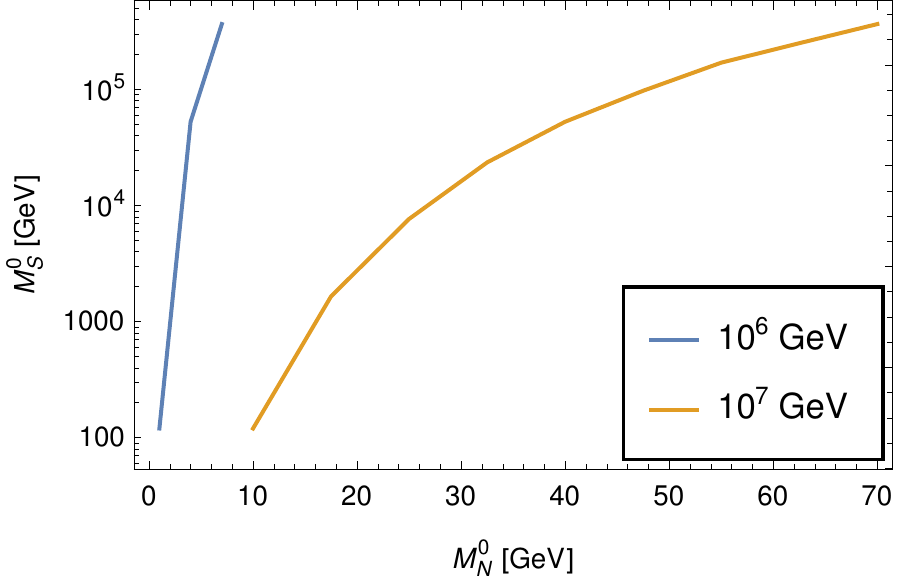}
    \caption{Scalar zero-temperature mass versus sterile neutrino zero-temperature mass. Under the assumption of a value for $v_S^0$, successful leptogenesis limits the two masses above the corresponding colored line.}
    \label{fig:mass-mass}
\end{figure}

\paragraph*{Implications of successful leptogenesis:}
We have seen that the dominant effect of additional scalars, in the parameter ranges discussed, is increased washout, such that successful leptogenesis imposes strong constraints on the possible values of Yukawa couplings $Y$, scalar self couplings $\lambda_S$, and the vev $v_S^0$.
Parameters that do not interfere with the leptogenesis mechanism are tiny Yukawa couplings with $Y\leq {\cal O}(10^{-7})$ and/or large vevs with $v_S^0 \geq {\cal O}(10^8)$~GeV, which corresponds to the decoupling limit.
Conversely, combinations of parameters where $Y$ and $v_S^0$ are both very small lead to very small $M_N^0$ and thus to oscillations that are too slow to generate an appreciable asymmetry before Sphaleron freeze out.

The domain with moderate vevs below the decoupling limit $10^6$~GeV~$\leq v_S^0 \leq 10^8$~GeV and Yukawa couplings $Y \geq 10^{-7}$ warrants further scrutiny.
In general also in this domain the generated asymmetry is reduced compared to the decoupling limit. For a fixed value of $v_S^0$ we can define the limiting value for $\lambda_S$ (or equivalently $M_S^0$)\footnote{We remark that the mass $M_S^0$ is obtained from diagonalising the scalar mass matrix, which includes an off-diagonal entry proportional to $\lambda_{SH} v_S^0 v_{EW}$. The condition that $S$ is thermalised, $\lambda_{SH} \geq {\cal O}(10^{-4})$ then gives a lower limit for $M_S^0$ for $\lambda_S < v_{EW}/v_S^0 \lambda_{SH}$.} where the generated asymmetry is half that of the asymmetry in the decoupling limit.
This allows us to define a minimum scalar mass for each sterile neutrino mass, for which leptogenesis is successful.
The resulting limits are shown in \cref{fig:mass-mass}, where the colored lines correspond to different values for $v_S^0$.

This figure can be intrepreted as follows. If scalar particles and sterile neutrinos are discovered, and their masses are below one of the colored lines, the corresponding vev has to be larger, or leptogenesis is not successful. As an example, consider $M_S^0=270$~GeV as motivated by the LHC multi-lepton anomalies \cite{vonBuddenbrock:2016rmr,vonBuddenbrock:2017gvy}. Our findings imply that the corresponding $M_N^0$ has to be smaller than $1$~GeV or $12$~GeV if $v_S^0=10^6$ or $v_S^0=10^7$, respectively. If $N$ with larger masses are discovered, this implies $v_S^0 \geq 10^8$~GeV, or that the BAU has to be generated in another way.

\paragraph*{Generalisation to multiple scalars:}
Here we considered the extension of the SM with sterile neutrinos and a single scalar field.
In scenarios where the SM is extended with sterile neutrinos and $n$ scalar singlet fields the sterile neutrinos can be even more connected to the thermal plasma, the zero-temperature masses of the sterile neutrinos and the sterile neutrino source terms are given by, respectively,
\begin{equation}
    M_N^0 = \sum_i Y_i v_{S_i}^0\,, \qquad \Gamma_{S_i} = \frac{Y_i\cdot Y_i}{16 \pi} \frac{1}{\rho^{eq}(x)} \frac{M_{S_i}(z)^2}{T_{EW}}\frac{z}{x^2} \int_{y_{0_i}}^\infty n_{s}(y) \dd y\,,
\end{equation}
where $Y_i$ and $v_{S_i}$ are the Yukawa coupling and vev of the scalar $S_i$. 
The $\Gamma_{S_i}$ are relevant for our discussion if and only if $S_i$ is thermalised and its mass $M_{S_i}^0$ is comparable to the oscillation time, $T_{EW}/z_{osc}$.

This brings the additional degree of freedom to increase the zero-temperature mass of the sterile neutrinos without increasing the washout, if a dominant contribution stems from a non-thermalised or very heavy scalar. This is comparable to allowing for Majorana mass terms.
In general we expect that in these scenarios the resulting asymmetry will be reduced by additional washout.

\paragraph*{Enhanced asymmetry production:}
A limited enhancement of the produced BAU is found for extremely small values of $\lambda_S$, relatively small values of $Y$ and large values of $v_S^0$, i.e.\ inside the blue area in fig.~\cref{fig:parameters}. 
The enhancement seems to occur when the timescales of scalar decays and sterile neutrino oscillations coincide. As an explicit example we discuss benchmark point E, see \cref{tab:points}, for this point the BAU is enhanced by about $10\%$ compared to the decoupling limit; from respectively $2.34 \times 10^{-10}$ in the decoupling limit to $2.5 \times 10^{-10}$ for benchmark point E.
This enhancement occurs for rather large $R_N$ production ($R_{N_2} \sim \mathcal{O}(1)$ at $z=1$). Thus, for a proper calculation of the BAU, inverse decay processes should be taken into account. As discussed before, this will reduce the predicted enhancement, according to our estimates the enhancement is in fact almost completely removed. We consider it unlikely that the enhancement will increase significantly in a full treatment when for example other momentum modes or inverse decays are taken into account properly.

Furthermore, we noticed that the sterile neutrino thermal mass (cf.~\cref{eq:Nmass}) increases the oscillations in the sterile sector, which in turn enhances the asymmetry production in the active one. 
However, in regions of parameter space where this effect is relevant, it is overcompensated by the enhanced washout from scalar decays.
If these two effects could be separated, a significant enhancement of the asymmetry production would be possible.
One way of separating these two effects is to have the time of $S$ symmetry breaking after the onset of oscillations, $z_S>z_{osc}$.
Parameters that realise this are denoted by the green area in \cref{fig:parameters}. However, they all lead to thermalisation of $N$.

In general, the asymmetry production is enhanced when the sterile neutrinos are more degenerate in mass.
However, the asymmetry production can also be enhanced without strong mass degeneracy when three flavors of sterile neutrinos are considered, as discussed in e.g.~Ref.~\cite{Abada:2018oly}. 

Another way that allows to separate zero-temperature sterile neutrino mass, finite temperature sterile neutrino oscillations, and the scalar decay into sterile neutrinos is given by a combination of thermalised and non-thermalised scalars, as discussed above. However, this goes beyond the scope of this work.

\paragraph*{Time of scalar symmetry breaking:}
For our numerical evaluation we have set the time scale $T_S$ at which the $S$ symmetry breaks and $v_S(T) \simeq v_S^0$ equal to the vev itself: $T_S = v_S^0$.
The time of symmetry breaking can be evaluated analytically if the field content of the theory is fixed, as is done for the case of the SM, for instance in Ref.~\cite{Dine:1992vs}.
From these arguments we expect that the time of symmetry breaking is proportional to 
\begin{equation}
T_S \propto \frac{v_S^0}{\sqrt{\lambda_S}}
\end{equation}
while the proportionality factors involve ratios of masses of possible additional field content. It is worth pointing out that, in the case of the SM, the energy scale of the symmetry breaking time $1/T_{EW} < v_{EW}$.

For our numerical evaluations we find that the exact time of symmetry breaking is irrelevant, as long as it occurs before the relevant time scales of leptogenesis.
In particular, symmetry breaking has to occur before the oscillations, which take place typically at $T_{osc} = {\cal O}(0.01) T_{EW}$. Therefore the corresponding energy scale of $T_S > 10^4$~GeV is sufficient not to introduce numerical effects on the asymmetry calculation.

\section{Conclusions}
Sterile neutrinos are well motivated from the light neutrino oscillations and they have been shown to successfully explain the Baryon Asymmetry of the Universe (BAU) through so-called ARS leptogenesis. 
Sterile neutrinos can be added in theories that include also other new fields, such as scalar bosons, which brings about the possibility of further interactions between the sterile neutrinos and the SM.

In this paper we considered an extension of the SM with two sterile neutrinos and one scalar singlet field in order to study the robustness of the ARS leptogenesis mechanism with respect to scalar extensions.
We took into account constraints from the light neutrino parameters and also discussed limits on the scalar sector from LHC searches.
We investigated the effect that the thermalised scalar has on the ARS leptogenesis mechanism.

We found that in our model the BAU of the $\nu$MSM is reproduced when the vev is at least as large as ${\cal O}(10^8)$~GeV and the Yukawa and scalar self couplings are at most of ${\cal O}(10^{-6})$, which we refer to as the decoupling limit.
In most of the remaining parameter space the thermalised scalar leads to enhanced sterile neutrino production at early times, resulting in a reduction of the predicted BAU compared to the decoupling limit.
A small enhancement of the BAU of ${\cal O}(10\% )$ is present for parameters close to the decoupling limit, i.e.\ $v_S^0\sim 10^8$~GeV and for scalar and heavy neutrino masses around and below the weak scale, respectively.

Our results are general for models with scalar singlets and with extended gauge sectors, provided that the additional field content does not thermalise the sterile neutrinos at any point of the Universe's history.
They can also be generalised to models with more than one scalar field, in which case the sterile neutrino zero-temperature mass and the scalar decay rate are sums over the scalar field content. In such models the zero-temperature sterile neutrino mass could be dominated by a scalar that is not thermalised, such that the Yukawa couplings in the sterile neutrino masses can be different from those in the decay rate of the thermalised scalar.

Our results can be useful when sterile neutrinos and scalar particles are discovered in the laboratory, such that their masses and the Yukawa coupling are known. In this case it is possible to infer whether or not the ARS mechanism is a valid possibility to create the BAU, or if another mechanism has to be invoked.

\bibliographystyle{unsrt}
\bibliography{references.bib}

\end{document}